\documentclass{article}
\usepackage{graphicx} 
\usepackage{latexsym}
\usepackage{enumerate}
\usepackage{bbm}
\usepackage{multirow}
\usepackage[small,bf]{caption2}
\usepackage{graphics}
\usepackage{epsfig}
\usepackage{booktabs}
\usepackage[textwidth=8em,textsize=small]{todonotes}
\usepackage{amsmath}
\usepackage{array}
\usepackage{amssymb}
\usepackage{amsthm}
\usepackage{amsfonts}
\usepackage{amsopn}
\usepackage{url}
\usepackage{xcolor}
\usepackage{colortbl}
\usepackage{hyperref}
\usepackage{cleveref}
\usepackage{appendix}
\newtheorem{theorem}{Theorem}[section]

\newtheorem{lemma}{Lemma}[section]

\theoremstyle{remark}\newtheorem{algorithm}{Algorithm}[section]
\usepackage{natbib}
\newcommand{\PP}{\mathbb P}

\newcommand{\bdelta}{{\boldsymbol{\delta}}}
\newcommand{\bY}{{\boldsymbol{Y}}}
\newcommand{\by}{{\boldsymbol{y}}}

\newcommand{\bOne}{\ensuremath{{\mathbf{1}}}}
\newcommand{\bZ}{{\boldsymbol{Z}}}
\newcommand{\bW}{{\boldsymbol{W}}}
\newcommand{\bz}{{\boldsymbol{z}}}

\newcommand{\obsbY}{{\boldsymbol{Y}^{\mbox{\scriptsize obs}}}}

\newcommand{\bv}{\boldsymbol{v}}
\newcommand{\bn}{\boldsymbol{n}}

\newcommand{\bbn}{\bar{\bar{\boldsymbol{n}}}}

\hypersetup{hidelinks}
\newcommand{\beq}{\begin{equation}}

\newcommand{\eeq}{\end{equation}}

\title{Exact and Conservative Inference for the Average Treatment Effect in Stratified Experiments with Binary Outcomes}
\author{Jiaxun Li\thanks{Department of Statistics, University of Michigan } \and Jacob Spertus\thanks{Department of Statistics, University of California, Berkeley } \and Philip B.\ Stark\footnotemark[2]}

\begin{document}

\maketitle

\begin{abstract}
    We extend methods for finite-sample inference about the average
    treatment effect (ATE) in randomized experiments with binary outcomes to 
    accommodate stratification (blocking).
    We present three valid methods that differ in their computational and statistical efficiency.
    The first method constructs conservative, Bonferroni-adjusted confidence intervals separately for the mean response in the treatment and control groups in each stratum, then takes appropriate weighted differences of their endpoints to find a confidence interval for the ATE. 
    The second method inverts permutation tests for the overall ATE, maximizing the $P$-value over all ways a given ATE can be attained. 
    The third method applies permutation tests 
    for the ATE in separate strata, then combines those tests to form a confidence interval for the
    overall ATE.
    We compare the statistical and computational performance of the methods using simulations and a case study.
    The second approach is most efficient statistically in the simulations, but a naive implementation requires \(O(\Pi_{k=1}^{K} n_{k}^{4})\) permutation tests, the highest computational burden among the three methods.
    That computational burden can be reduced to \(O(\sum_{k=1}^K n_k \times\Pi_{k=1}^{K} n_{k}^{2})\) 
    if all strata are balanced and to \(O(\Pi_{k=1}^{K} n_{k}^{3})\) otherwise.
\end{abstract}

\section{Introduction}

In a randomized binary experiment with binary outcomes, $n$ subjects are randomized into a 
``treatment'' group of size $m$ and a ``control'' group of size $n - m$.
For each subject, a binary response is measured, e.g., whether the subject survives to time $t$.
Such \emph{randomized controlled trials} (RCTs) with binary treatments and binary outcomes have been studied at least since Fisher's seminal work \citep{fisher35}.
\textit{Stratification} (or \textit{blocking}) is a widely used experimental design that can increase the statistical efficiency and logistical feasibility of RCTs \citep{imbens2015causal}.
Our paper develops methods for exact inference in stratified RCTs
with binary treatments and binary outcomes.

\subsection{Potential outcomes and average treatment effects}

In many situations, a reasonable model for causal inference in randomized experiments characterizes each subject's
responses using
\emph{potential outcomes}:
$y_j(0)$ is the response that subject $j$ would have if assigned to control, and $y_j(1)$ is the response subject $j$ would have if assigned to treatment
\citep{neyman1923,rubin74}.
The numbers $\{y_j(k)\}_{j=1}^{n}$, $k = 0, 1$, are considered fixed before the randomization.
If subject $j$ is assigned to control, we observe $y_j(0)$; if subject $j$ is assigned to treatment, we observe $y_j(1)$.
We do not observe both $y_j(0)$ and $y_j(1)$ for any subject.
Implicit in this model is \emph{non-interference}:\footnote{%
    Modeling the treatment as binary also assumes there is \textit{no hidden variation in treatment}. 
    Together these comprise the stable unit treatment value assumption (SUTVA). See, e.g., \S 1.6 of \citet{imbens2015causal}.
} 
the observed response of subject $j$ depends only on whether subject $j$ is assigned to treatment or to control, and not
on the assignment of any other subjects.\footnote{%
    This would not be a good assumption in some circumstances, for instance, studying the effect of vaccination on the propagation of a communicable disease.
}

Let $\by := ((y_j(1), y_j(0)))_{j=1}^n$ 
be the matrix of potential outcomes for the $n$ subjects.
The \emph{average treatment effect} (ATE) $\tau$ is the mean response if all subjects were assigned to treatment minus the mean response if all subjects were assigned to control:
$$
\tau(\by) := \frac{1}{n} 
\sum_{j=1}^n y_j(1) -
\frac{1}{n} \sum_{j=1}^n y_j(0).
$$
The ATE is the primary quantity of interest in many RCTs.\footnote{%
    There was a longstanding disagreement between
    Fisher and Neyman about the ``correct'' null hypothesis
    to test
    \citep{feinbergTanur96,wuDing21}.
    Fisher advocated testing the 
    ``strong'' null
    that treatment has no effect whatsoever on any individual, i.e., that $y_j(0) = y_j(1)$ for all $j$.
    Neyman advocated testing 
    the ``weak'' null hypothesis that $\tau = 0$.
}
The ATE is not observable because it involves both potential outcomes for every subject, but only one potential outcome is observed.
Many applications involve tests and confidence sets for the ATE in binary experiments with binary outcomes, including agricultural experiments \citep{ReboudEtal2009}, medical experiments \citep{leonEtal2010}, marketing \citep{sahni2013}, and others. 
It is common to use tests and confidence sets for the ATE based on asymptotic theory, which can be anti-conservative
\citep{rigdon_binary_2015,li_binary_2016}.

\subsection{Stratified random experiments}

A number of methods have been proposed for making exact or conservative inferences about the ATE in unstratified experiments
with binary treatments and binary outcomes
\citep{santer_2by2_1980,branson_bernoulli_2019,chiba_binary_2015,rigdon_binary_2015,li_binary_2016, aronow_fast_2023}.

In a stratified experiment, the population of subjects is partitioned into $K$ disjoint
strata.
Subjects are randomized to treatment or control within each stratum by simple random sampling, independently across strata.
Stratification is common in clinical trials, where strata may comprise subjects recruited at a particular center
(often stratified further by gender and health covariates); indeed,
\citet{bruce_RCT_2022} estimates that almost two-thirds of clinical trials use some form of stratification.

To the best of our knowledge, data from stratified experiments are generally analyzed using asymptotic tests and confidence sets,
including methods based on regression
\citep{imbens2015causal}.
The only proposed exact or conservative methods for inference about the ATE from stratified binary experiments we know of are those of \citet{rigdon_binary_2015} and \citet{chiba_strata_2017}.
\citet{rigdon_binary_2015} sketches an approach based on permutation tests but dismisses it as computationally intractable.
\citet{chiba_strata_2017} proposes an exact approach and also concludes it is impractical when the number of subjects or strata is large.

This paper makes three contributions:
\begin{enumerate}[i.]
    \item It develops 
computationally tractable methods for testing hypotheses about the ATE and forming confidence intervals for the ATE from stratified binary experiments with binary outcomes.
    \item It improves the computational efficiency of some extant methods.
    \item It compares the statistical and computational efficiency of a variety of methods using simulations.
\end{enumerate}
The methods are illustrated using data from a clinical trial of vedolizumab versus placebo for chronic pouchitis, stratified by baseline antibiotic use. 

\section{Notation}

We generally use Latin letters from the middle of the alphabet, such as $K$, $n$, $m$, $i$, $j$, and $k$, to denote nonnegative integers.
If $K$ is a positive integer, $[K] := \{1, \ldots, K\}$.
Uppercase Latin letters from the end of the alphabet, such as $Y$ and $Z$, generally denote random variables.
We generally use bold font to denote tuples, vectors, and matrices, e.g., $\bv$, $\bn$, $\bZ$.

A population of $n$ subjects is partitioned into $K$ strata. 
Stratum $k$ contains $n_k$ subjects; simple random sampling assigns
$m_k$ of them
to active treatment and
$n_k-m_k$ to control.
Assignments are independent across strata.
The total number of subjects assigned to treatment
is $m := \sum_{k=1}^{K} m_k$.
Because the strata partition the population, $n = \sum_{k=1}^{K} n_k$. 
The potential outcomes for the $j$th subject in the $k$th stratum are 
$$\by_{kj} := (y_{kj}(1), y_{kj}(0)) \in \{0,1\}^2.$$
Let $\by_k := (\by_{kj})_{j=1}^{n_k}$ be the potential outcomes for the subjects in the $k$th stratum, and let
$\by := (\by_k)_{k=1}^{K}$ denote the entire collection of individual potential outcomes. 
The ATE is
\[
  \tau(\by) := \frac{1}{n}\sum_{k=1}^{K} \sum_{j=1}^{n_k} [y_{kj}(1)-y_{kj}(0)].
\]
Let $Z_{kj} = 0$ if the $j$th subject in the $k$th stratum is assigned to control and $Z_{kj} = 1$ otherwise, and 
let $\bZ_k := (Z_{kj})_{j=1}^{n_j}$.
The \emph{treatment assignment table} is
$\bZ := (\bZ_k)_{k=1}^{K}$, a random ragged binary table.
The outcome for the $j$th subject in the $k$th stratum is 
$$
Y_{kj} = Z_{kj}y_{kj}(1)+(1-Z_{kj})y_{kj}(0).
$$
The vector of outcomes for the $k$th stratum is
$\bY_k := (Y_{kj})_{j=1}^{n_k}$
and the entire collection of outcomes is 
$\bY := (\bY_k)_{k=1}^{K}$.
The \textit{ATE in stratum $k$} is
\[
    \tau_k(\by) := \frac{1}{n_k} \sum_{j=1}^{n_k}[y_{kj}(1)-y_{kj}(0)].
\]
An unbiased estimate of $\tau_k(\by)$ is
\[
    \hat{\tau}_k(\bY,\bZ) := 
    \frac{1}{m_k} \sum_{j=1}^{n_k} Z_{kj}Y_{kj} 
    - 
    \frac{1}{n_k-m_k}\sum_{j=1}^{n_k}(1-Z_{kj})Y_{kj}.
\]
The ATE can be written $\tau(\by) = \frac{1}{n}\sum_{k=1}^{K} n_k \tau_k(\by)$, and its usual unbiased estimator is
\begin{align*}
    \hat{\tau}(\bY, \bZ) :=& 
     \frac{1}{n} \sum_{k=1}^{K}n_k 
    \left[
        \frac{1}{m_k} \sum_{j=1}^{n_k}Z_{kj}Y_{kj} - \frac{1}{n_k-m_k} \sum_{j=1}^{n_k}(1-Z_{kj})Y_{kj}
    \right] \\
    =& \frac{1}{n} \sum_{k=1}^{K} n_k \hat{\tau}_k(\bY, \bZ),
\end{align*}
a weighted average of the within-stratum ATE estimates. 

The number of subjects in stratum $k$ whose response if assigned to treatment is $a\in\{0,1\}$ and whose response if assigned to control is $b\in\{0,1\}$ is 
\begin{equation}
    v_{kab} := \sum_{j=1}^{n_k} \bOne \{y_{kj}(1)=a, y_{kj}(0)=b\}.
\end{equation}
Let $\bv_k :=(v_{k11}, v_{k10}, v_{k01}, v_{k00})$. 
The
\emph{potential outcome table} $\bv$ summarizes the individual potential outcomes using $2 \times 2 \times K$ integers:
$\bv := (\bv_k)_{k=1}^{K}$.
The stratumwise ATEs and the ATE can be written as functions of $\bv$:
\begin{equation}
    \tau_k(\bv) = \frac{1}{n_k} (v_{k10}-v_{k01}), \quad k = 1, 2, \ldots, K;
    \label{tau_k}
\end{equation}
\begin{equation}
\label{eqn:ate_potential_outcomes}
    \tau(\bv) = \frac{1}{n} \sum_{k=1}^{K} (v_{k10}-v_{k01}).
\end{equation}
The data also can be summarized by a table of integers:
the number of subjects in stratum $k$ whose treatment assignment is $u$ and whose observed response is $w$ is 
\begin{equation}   
    n_{kuw} := \sum_{j=1}^{n_k} \bOne
    \{Z_{kj}=u, Y_{kj}=w\}.
\end{equation}
Let $\bn_k :=(n_{k11}, n_{k10}, n_{k01}, n_{k00})$. 
The \emph{outcome table} is 
$\bn :=(\bn_k)_{k=1}^K$.
The estimated stratumwise ATE $\hat{\tau}_k$ and the estimated ATE $\hat{\tau}$ can be written as functions of $\bn$:
\begin{equation}
    \hat{\tau}_k(\bn) = \frac{1}{n_k} \left ( \frac{n_{k11}}{m_k} - \frac{n_{k01}}{n_k-m_k} \right ), \quad k= 1, 2, \ldots, K.
\label{tau_hat_k}
\end{equation}
\begin{equation}
    \hat{\tau}(\bn) = \frac{1}{n} \sum_{k=1}^{K} n_k \left ( \frac{n_{k11}}{m_k} - \frac{n_{k01}}{n_k-m_k} \right ). 
\label{tau_hat}
\end{equation}

Let $\bbn(\bv, \bz)$ denote the outcome table that would result from the treatment assignment table $\bZ=\bz$ applied to a ``canonical unpacking'' of the potential outcome table $\bv$ into a full set of potential outcomes for all $n$ subjects.\footnote{%
    Here is an example of a canonical unpacking:
    set
    $y_{kj}(1) := 1$ and $y_{kj}(0) := 1$ for the first $v_{k11}$ subjects in stratum $k$; 
    $y_{kj}(1) := 1$ and $y_{kj}(0) := 0$ for the next $v_{k10}$ subjects in stratum $k$; 
    $y_{kj}(1) := 0$ and $y_{kj}(0) := 1$ for the next $v_{k01}$ subjects in stratum $k$;
    and 
    $y_{kj}(1) := 0$ and $y_{kj}(0) := 0$ for the last $v_{k00}$ subjects in stratum $k$.
}
The outcome table $\bn$ constrains the potential outcome table $\bv$ algebraically:
$\bv$ is 
\textit{algebraically compatible} with $\bn$ if 
there is some treatment assignment table $\bz$ for which $\bn = \bbn(\bv, \bz)$ and $\PP(\bZ=\bz)>0$.

\section{Inferences about the ATE in unstratified experiments}
\label{sec:prework}

Suppose there is only one stratum:
subjects are assigned to treatment by simple random sampling.
We suppress the stratum subscript in this section.
For example,
$\bv :=(v_{11}, v_{10}, v_{01}, v_{00})$ 
denotes the potential outcome table and 
$\bn :=(n_{11}, n_{10}, n_{01}, n_{00})$ denotes the observed outcome table.

\citet{rigdon_binary_2015} and \citet{li_binary_2016}
present two approaches to finding confidence intervals for the ATE in 
this problem.
One is based on hypergeometric confidence sets; the other is based on inverting permutation tests.

\subsection{Combining hypergeometric confidence bounds} 
\label{sec:hyper_unstratified}
Let $v_{1 \bullet} := v_{10}+v_{11}$ be the number of subjects whose response would be 1 if assigned to the active treatment and $v_{\bullet 1} := v_{01}+v_{11}$ be the number of subjects whose response would be 1 if assigned to control. 
The ATE can be written
\beq 
    \tau(\bv) := (v_{10}-v_{01})/n=(v_{1 \bullet}-v_{\bullet 1})/n.
\eeq
One can construct a conservative confidence interval for the ATE 
by combining simultaneous confidence intervals for $v_{1 \bullet}$ and $v_{\bullet 1}$.
When there is only one stratum, the treatment and control groups are simple random samples of the $n$ subjects,
so
\[
    n_{11} \sim \text{Hypergeo}(v_{1 \bullet}, n, m) \mbox{ and } n_{01} \sim \text{Hypergeo}(v_{\bullet 1}, n, n-m).
\]
Standard methods can be used to find 
$1 - \alpha/2$ confidence intervals for the
hypergeometric parameters $v_{1 \bullet}$ and $v_{\bullet 1}$.
Bonferroni adjustment (i.e., using confidence level $1-\alpha/2$ rather than $1-\alpha$) ensures that the pair of intervals has simultaneous confidence level at least $1-\alpha$.
Let $v_{1 \bullet}^{\ell}$ be the lower confidence bound for $v_{1 \bullet}$;
let 
$v_{1 \bullet}^{u}$ be the upper confidence bound for $v_{1 \bullet}$; and define 
$v_{\bullet 1}^{\ell}$ and $v_{\bullet 1}^{u}$
analogously.
Then
$$
    [(v_{1 \bullet}^{\ell} - v_{\bullet 1}^{u})/n,(v_{1 \bullet}^{u}
    - v_{\bullet 1}^{\ell})/n]
$$ 
is a conservative 
$1-\alpha$ confidence interval for $\tau$.
The method proposed
in section~\ref{sec:exact_ci}
extends this strategy to stratified  experiments.
Other methods for finding confidence intervals for the ATE using simultaneous hypergeometric confidence intervals are given in \citet{rigdon_binary_2015,li_binary_2016}.
This approach to constructing a confidence interval for the ATE is computationally inexpensive but
unnecessarily conservative because of Bonferroni adjustment; we now consider 
sharper approaches.

\subsection{Inverting hypothesis tests}
\label{sec:inverting_unstratified}

One can construct confidence intervals for the ATE
using the general duality between confidence sets and hypothesis tests.
Consider the null hypothesis
\[
    H_{0}(\bdelta) : y_i(1) - y_i(0) = \delta_i, \;\; i = 1, \ldots, n,
\]
where $\bdelta = (\delta_1, \delta_2, \ldots, \delta_n)$ is a known vector. 
After $\bY$ and $\bZ$ have
been observed, this yields a \emph{simple} or \emph{sharp} null \citep{rubin_comment_1980}:
$\bdelta$, $\bY$, and $\bZ$
are consistent with only one full set of potential outcomes $\by$, which determines the randomization distribution of any test statistic $T(\bY, \bZ)$.

Suppose that larger values of the test statistic $T$
are considered to be stronger evidence against the null.
Then a natural way to define a $P$-value for the hypothesis
that the full potential outcome table is $\by$
(when $\bZ=\bz$ and $\bY=\obsbY$) is
\begin{equation} \label{eq:p-value-def}
    \PP_\by \left \{ T(\obsbY, \bW) \ge T(\obsbY, \bz )  \right \},
\end{equation}
where the probability on the right hand side is calculated with respect to the random assignment $\bW \sim \bZ$. 

Following  
\citet{rigdon_binary_2015,li_binary_2016,aronow_fast_2023}, 
take 
\begin{equation} 
\label{eq:T-def}
T(\bY, \bZ) := | \hat{\tau}(\bY, \bZ) - \tau(\by) |.
\end{equation}
(There are other reasonable choices
for $T$, but this test statistic works well in practice; see \Cref{sec:other_test_stat}.)
Recall that the observed outcomes can be summarized by
the outcome table $\bn$. 
Under $H_0(\bdelta)$, there is a unique potential outcome table $\bv_{\bdelta}$ for which the treatment effects are $\bdelta$ and $\bbn(\bv_{\bdelta}, \bz) = \bn$.
For the test statistic $T$ in (\ref{eq:T-def}), the probability (\ref{eq:p-value-def}) is:
\begin{equation}
    \PP_{\by} \left \{ |\tau(\bv_\bdelta)-\hat{\tau}(\bbn(\bv_\bdelta,\bZ))| 
    \ge |\tau(\bv_\bdelta) - \hat{\tau}(\bn)| \right \}.
    \label{eq: p-value-def-strata}
\end{equation}
We reject $H_0(\bdelta)$ if that probability does not exceed $\alpha$.
This procedure is often called the \emph{Fisher randomization test} (FRT) or \emph{permutation test}
of the sharp null $H_0(\bdelta)$.

We can obtain a confidence interval
for the ATE by inverting a collection of permutation tests:
for each possible value of $\bdelta$, we perform a permutation test of $H_0(\bdelta)$.
The $1-\alpha$ confidence set for the ATE comprises all values of $\frac{1}{n}\sum_{i=1}^n \delta_i$ among treatment effects $\bdelta$ for which $H_0(\bdelta)$ is not rejected.
In practice, a confidence interval may be more useful than a possibly disconnected confidence set, so we focus on finding the smallest and largest and smallest values of $\tau$ in the confidence set. 

Since the size of the support of $\bZ$ is $\binom{n}{m}$, it is 
computationally intractable to compute the
``full randomization'' $P$-value when $n$ is large and $m$ is not close to $0$ or $n$.
Instead, we use an
exact Monte Carlo $P$-value for a related randomized test:
we draw $R$ treatment allocations $\{\bW_r\}_{r=1}^R$
at random, uniformly,
according to the experimental design,
creating $R$ sets of observed responses
for a given hypothesized potential outcome table.
Then 
\begin{equation} \label{eq:p-value-def-random}
p(\bv_\bdelta, \bn) :=
\frac{\# \left \{r \in [R] : 
|\tau(\bv_\bdelta)-\hat{\tau}(\bbn(\bv_\bdelta,\bW_r))| 
    \ge |\tau(\bv_\bdelta) - \hat{\tau}(\bn)| \right \} + 1}{R+1}
\end{equation}
is an exact $P$-value for a randomized test of the hypothesis $H_0(\bdelta)$
\citep{dwass57,ramdasEtal23,glazerStark25}.
One may take $R$ to be arbitrarily small to reduce the computational burden, but as $R$ shrinks, the smallest attainable $P$-value grows (it is $1/(R+1)$).

While the null hypothesis for the permutation test specifies the individual treatment effects, the permutation distribution of the test statistic is fully determined by the potential outcome table $\bv$. 
Therefore, the number of permutation tests that need to be performed is at most the number of potential outcome tables tested; moreover, the same set of $R$ allocations can be used to test every table to reduce the computational burden
\citep{glazerStark25}.
There are at most $\binom{n+3}{3}$ potential outcome tables $\bv=(v_{11},v_{10},v_{01},v_{00})$ with $v_{11}+v_{10}+v_{01}+v_{00}=v$. 
Hence, the computational complexity of the overall confidence procedure is at most $O(n^3)$. 
\citet{li_binary_2016} showed that the number of potential outcome tables that must be considered is at most $O(n^2)$.
\citet{aronow_fast_2023} showed that
for balanced experiments, testing $O(n \log n)$ potential outcome tables
suffices. 
Section~\ref{sec:reduce_time} extends their methods to stratified experiments. 


\section{Inferences about the ATE in Stratified Experiments}
\label{sec:exact_ci}
This section extends  Section~\ref{sec:prework} to stratified experiments.

\subsection{Extending the hypergeometric confidence bound approach to stratified experiments}
\label{sec:hyper_stratified}
We describe a procedure analogous to that outlined in \citet[Section~5]{rigdon_binary_2015}.
For each stratum $k$, let $v_{k1\bullet} := v_{k10}+v_{k11}$ 
denote the number of subjects in stratum $k$ whose response would be 1 if assigned to active treatment, and let
$v_{k \bullet1} := v_{k01}+v_{k11}$ denote the number of subjects in stratum $k$ whose response would be 1 if assigned to control. 
Define $v_{1\bullet} := \sum_{k=1}^{K} v_{k1\bullet}$ and $v_{\bullet 1} := \sum_{k=1}^{K} v_{k\bullet 1}$.
Then the ATE can be written
$\tau = (v_{1\bullet}-v_{\bullet 1})/n$. 
We can obtain a conservative confidence interval for $\tau$ in the stratified case using a method analogous to that in Section~\ref{sec:hyper_unstratified}
by combining simultaneous confidence intervals for $v_{1 \bullet}$ and $v_{\bullet 1}$
from stratified samples.
Specifically, suppose that
$[v^\ell_{1\bullet}, v^u_{1 \bullet}]$ and
$[v^\ell_{\bullet 1}, v^u_{\bullet 1}]$ are simultaneous $1-\alpha$ confidence intervals for $v_{1\bullet}$ and $v_{\bullet 1}$, respectively.
Then 
\begin{equation}
    \left [ \frac{v^\ell_{1\bullet} - v^u_{\bullet 1}}{n}, \frac{v^u_{1\bullet}-v^\ell_{\bullet 1}}{n} \right ]
\end{equation}
is a conservative $1-\alpha$ confidence 
interval for the ATE $\tau$.

Simultaneous confidence intervals for 
$v_{1\bullet}$ and $v_{\bullet 1}$
can be constructed from stratified
experimental data in a variety of ways.
We consider two: the method of \citet{wendell_strata_1996} with the Bonferroni
adjustment for simultaneity,
and the ``efficient stratified inference'' method 
of \citet{stark19} with
the Bonferroni adjustment.
The method of \citet{wendell_strata_1996}
(W-S, henceforth) is computationally intractable when the number of strata is greater than about 3;
the method of \citet{stark19} (ESI, henceforth) is still computationally inexpensive
when there are many strata.

\subsection{Stratified permutation tests}
\label{sec:invert_permute}
When the experiment is stratified, the approach in Section~\ref{sec:inverting_unstratified} can still be used if the randomization is modified to match the stratification.
Consider the hypothesis
\begin{equation}
    H_0(\bdelta):y_{kj}(1)-y_{kj}(0)=\delta_{kj}, \;  
    k = 1, \ldots, K; \; 
    j = 1, \ldots, n_k,
    \label{h0_stratified}
\end{equation}
where 
$$
\bdelta=(\delta_{11}, \delta_{12},\ldots,\delta_{1n_1},\delta_{21},\delta_{22},\ldots,\delta_{2n_2} \ldots\ldots,\delta_{K1},\delta_{K2}\ldots,\delta_{Kn_K})
$$ 
is a known vector. 
The values of $\bdelta$ and $\bZ$ fully determine the potential outcomes, which can be summarized in a 
4~by~$K$ table $\bv_\bdelta =(\bv_{\bdelta, k})_{k=1}^{K}$. 
In a stratified randomized experiment, $\bZ$ is uniformly distributed over the set
\begin{equation} 
 \big \{
 ((z_{kj})_{j=1}^{n_k})_{k=1}^K 
: z_{kj} \in \{0, 1\}, ~\forall~ (k, j) \mbox{ and }
\sum_{j=1}^{n_k} z_{kj}=m_k,
 \quad k=1, 2, \ldots , K
 \big \}.
\end{equation}
A $P$-value for $H_0(\bdelta)$ is
given by \Cref{eq:p-value-def-random}.
To find a confidence interval for the ATE, we can in principle perform a permutation test of $H_0(\bdelta)$ for each possible value of $\bdelta$. 
The largest and smallest values of $\frac{1}{n}\sum_{k=1}^K\sum_{i=1}^{n_k} \delta_{kj}$ among values of $\bdelta$ that are not rejected  are the upper and lower endpoints of the confidence interval. 

This method is related to
that of \citet{chiba_strata_2017}
and
of \citet[\S5]{rigdon_binary_2015}. 
As those papers
mention, the computational cost of this approach is high,
since for each stratum in the potential outcome table 
$\bv=(\bv_1,\ldots,\bv_K)$, 
there are $O(n_k^3)$ possible values of $\bv_k$ to consider. 
Roughly speaking, the total number of permutation tests is on the order of $O(n^{3K})$. 
However, if the allocation is balanced in $\ell$ of the $K$ strata, we can reduce the computational cost to $O(n^{3K-\ell+1})$. 
The algorithm and the proof that it works are in Appendix~\ref{sec:reduce_time}.

We call this method ``SPT'' for \emph{stratified permutation test}.
The next section introduces an approach that decouples the computations in different strata,  which reduces the total computational burden.

\subsection{Inverting stratified Permutation Tests that use Combining Functions}
\label{sec:invert_permute_combine}
This section extends the ideas in  \ref{sec:inverting_unstratified} and \ref{sec:hyper_unstratified}.
Consider the intersection of hypotheses 
about the ATE in each stratum:
\[
H_0(\tau_{10}, \ldots,\tau_{K0}):\{ \tau_{k} = \tau_{k0}, ~k = 1, \ldots, K\}.
\]
(Note that, as opposed to $H_0(\boldsymbol{\delta})$, this hypothesis is about the ATE in each stratum rather than about the treatment effect for each subject separately.)
We can test this intersection hypothesis
by finding a $P$-value for each  hypothesis $\tau_k = \tau_{k0}$, $k=1, \ldots, K$, then combining those $K$  stratumwise $P$-values to obtain a $P$-value for the intersection hypothesis. 
The $P$-value for the composite hypothesis $H_{k0}: \tau_k = \tau_{k0}$ in stratum $k$
is the largest $P$-value among all simple hypotheses (potential outcome tables for stratum $k$) that have ATE $\tau_{k0}$.
In other words, the $P$-value of the composite hypothesis is the maximum $P$-value over a multidimensional nuisance parameter, the potential outcome table in stratum $k$:
\[
p(\tau_{k0}) := \max_{\tau(\bv_k)=\tau_{k0}} p(\bv_k,\bn_k),
\]
where $p(\bv_k,\bn_k)$ is defined in (\ref{eq:p-value-def-random}).
To construct a $1-\alpha$ confidence set for $\tau$, we test the hypothesis $H_0(\tau_{10}, \ldots, \tau_{K0})$ at level $\alpha$ for each possible combination of $\tau_{10}, \ldots, \tau_{K0}$ and include $\sum_{k=1}^{K} n_k \tau_{k0}/n$ in the confidence set if $H_0(\tau_{10}, \ldots, \tau_{K0})$ is not rejected.

A conservative $P$-value  for $H_0(\tau_{10}, \ldots, \tau_{K0})$ 
can be found from the stratum $P$-values by using Fisher's combining function as the test statistic:
\[
t_F(\tau_{10},\ldots,\tau_{K0})=-2 \sum_{k=1}^{K}\ln p(\tau_{k0}).
\]
Under the intersection null, the distribution of $t_F(\tau_{10},\ldots,\tau_{K0})$
is dominated by the chi-square distribution with $2K$ degrees of freedom. 
Let $\chi_{2K}^2(\alpha)$ denote the $1-\alpha$ quantile of the chi-square distribution with $2K$ degrees of freedom. 
Then we can reject $H_0(\tau_{10}, \ldots, \tau_{k0})$ if $t_F(\tau_{10}, \ldots, \tau_{K0}) \ge \chi_{2K}^2(\alpha)$. 
Reasons to use Fisher's combining function
are discussed in Appendix~\ref{sec:combining_choice}.
We call this method ``CPT'' for \emph{combined permutation test}.

Unlike the SPT method described above, CPT does not require testing every combination of 
 stratumwise potential outcome tables $(\bv_1, \bv_2, \ldots, \bv_K)$. 
Instead, one can test in each stratum separately to obtain a $P$-value for each possible value of $\tau_k$
(i.e., $\tau_k$ in the range $(n_{k11}+n_{k00}-n_k)/n_k$ to $(n_{k01} + n_{k10})/n_k$)
and then combine those $P$-values
across strata.
At most $n_k$ values of $\tau_k$ are algebraically compatible with the data in stratum $k$, resulting in a time complexity of $O(\prod_{k=1}^{K} n_k)$ for finding the confidence interval
for $\tau$. 
The number of operations
to find a $P$-value for stratum $k$ is at most 
$O(R n_k^{3})$,
where $R$ is the number of Monte Carlo allocations.
\citet{aronow_fast_2023}
chose $R$ to be proportional
to $\log (n_k \log n_k)$, but
one may construct conservative randomized $P$-values for any $R$ \citep{dwass57,ramdasEtal23,glazerStark25}, which leads to 
an overall operation count 
 $O(\max\{\sum_k^K n_k^3, \prod_k^K n_k\}).$
This is much less than 
$O(\prod_{k=1}^K n_k^3)$ operations for the 
``global''
Monte Carlo method in Section~\ref{sec:invert_permute}.


\section{Illustration}
\subsection{Simulations}
This section compares the four new methods with approximate $1-\alpha$ Wald confidence intervals
based on the normal approximation:
\begin{equation}
   \label{equ_wald_ci}
    \hat{\tau}\pm z_{1-\alpha/2}
    \left \{
  \sum_{k=1}^{K} \left ( \frac{n_k}{n} \right )^2
        \left (
        \frac{\widehat S^2_k(1)}{m_k}+\frac{ \widehat S^2_k(0)}{n_k-m_k}
        \right )
    \right \}^{1/2},
\end{equation}
where 
$$
\widehat{S}^2_k(1) := (m_k-1)^{-1} \sum_{j=1}^{n_k}Z_{kj}(Y_{kj} -
\hat{\bar {Y_k}}(1)),
$$
$$
\widehat{S}^2_k(0) := (n_k-m_k-1)^{-1}\sum_{j=1}^{n_k}(1-Z_{kj})(Y_{kj} -\hat{\bar {Y}}_k(0)),
$$ 
$$
\hat{\bar {Y}}_k(1) := \frac{1}{m_k} \sum_{j=1}^{n_k} Z_{kj} Y_{kj},
$$ 
$$
\hat{\bar {Y}}_k(0) := \frac{1}{n_k-m_k} \sum_{j=1}^{n_k} (1-Z_{kj}) Y_{kj},
$$
and 
$z_{1-\alpha/2}$ is the $1-\alpha/2$ quantile of the standard normal distribution. 
Wald confidence intervals for the ATE in stratified experiments are asymptotically valid \citep{liuYang20},
but do not guarantee their nominal coverage in practice. 
Tables~\ref{tab_simu} and \ref{tab_simu_large} report the mean widths and empirical coverage rates of the confidence intervals for a variety of potential outcome tables.\footnote{%
The simulations in this paper used a ROG14 laptop with AMD Ryzen~7 6800HS CPU, 4.00 GHz clock, about 580~MB memory, Microsoft Windows~11 version 24H2, python version 3.10.11, and visual studio code version 1.85.}
Table~\ref{tab_simu} gives results for all methods on problems with up to 3 strata;
Table~\ref{tab_simu_large} reports results for 3, 4, and 5 strata for which two of the methods (SPT and W-S) did not complete within 30 minutes.
The simulations were conducted as follows:
\begin{enumerate}
    \item Set the number of strata $K$, the stratum sizes $\{n_k\}_{k=1}^K$, and the potential outcome table $\bv$ (which determines the ATE $\tau$).
    Set the sizes of the treated groups in each stratum $\{m_k\}$.
    For $r=1, \ldots, 100$:
    \begin{itemize}
        \item For $k = 1, \ldots, K$, randomly assign $m_k$ subjects  to treatment and generate the corresponding observed outcome table $\bn$.
        \item Compute the (nominal or actual) 95\% confidence interval using each of the five methods.
        (The Monte Carlo methods used $R=100$ random allocations.)
    \end{itemize}
    \item Report the average width, empirical coverage rate, and mean running time.
\end{enumerate}

Rows~2, 4, 5 and 12 of Table~\ref{tab_simu} and row~3 of Table~\ref{tab_simu_large} show that the Wald method does not always achieve its nominal coverage rate.
Among the exact/conservative methods, SPT generally produced the narrowest confidence intervals (when it completed), followed by
W-S (when it completed) and CPT, and finally ESI.
For larger problems, ESI sometimes beats CPT
in mean width; moreover, its execution time is far smaller (roughly 500 times smaller in one example with 5 strata).


There is an apparent trade-off between statistical efficiency and computational efficiency: both for
permutation tests and Bonferroni-adjusted population confidence bounds, the method that generally leads to the narrowest confidence bounds in simulations is computationally intractable when there are more than a few strata of moderate size.
In particular, Tables~\ref{tab_simu} and \ref{tab_simu_large} show that W-S and SPT are impractical when $K \ge 3$ and $n \ge 100$. 
Computing confidence intervals with SPT, which produced the narrowest confidence intervals on average, 
required up to 30~minutes for $K=3$ and $n=60$, but did not complete within 30~minutes in examples with 4 and 5 strata.
Even using the more efficient algorithms in Section~\ref{sec:reduce_time}, 
the approach is computationally intractable for problem sizes that arise frequently in real experiments.

\begin{table}
    \centering
    \small
    \resizebox{\columnwidth}{!}{%
    \begin{tabular}{p{2.1cm}p{1.7cm}p{2cm}|p{1cm}p{1cm}p{1cm}p{1cm}p{1cm}}
    \toprule
         $\bv$ & $\bn$ & $\tau$ &  \textbf{Wald} & \textbf{ESI} & \textbf{W-S} & \textbf{CPT} & \textbf{SPT}\\
         \hline
         
         [10, 10, 10, 10], \newline  [10, 10, 10, 10]& (10, 10) & (0, 0) &  0.51 \newline  96\% \newline 0s  & 0.67 \newline 100\% \newline 0.01s & 0.57 \newline 100\% \newline 12.78s &0.57 \newline 100\% \newline 0.46s  &  \textbf{0.50} \newline 99\% \newline 181.27s\\
  \hline       
        [3, 8, 4, 5], \newline  [0, 19, 1, 0] & (15, 15) & (0.2, 0.9) &
         0.53 \newline  \textit{ 90\% }\newline 0s  & 0.75 \newline 98\% \newline 0s & 0.64 \newline 100\% \newline 0.9s &0.63 \newline 100\% \newline 0.05s  & \textbf{0.54} \newline 100\% \newline  3.97s\\
  \hline       

         [3, 23, 2, 2], \newline  [4, 2, 30, 4] & (5, 30) & (0.7, -0.7) & 
         0.40 \newline  99\% \newline 0s  & 0.70 \newline 100\% \newline 0.01s & 0.68 \newline 100\% \newline  6.1s& 0.61 \newline 100\% \newline 0.19s  & \textbf{0.54} \newline 100\% \newline 21.78s\\
  \hline       

         [2, 24, 0, 4], \newline  [1, 26, 2, 1] & (5, 25)& (0.8, 0.8) & 
         0.38 \newline  \textit{65\%} \newline 0s  & 0.72 \newline 100\% \newline 0.01s & 0.62 \newline 100\% \newline 0.82s& 0.53 \newline 100\% \newline 0.08s  & \textbf{0.43} \newline 99\% \newline 17.54s\\
  \hline       

         [1, 0, 9, 0], \newline  [0, 40, 0, 0] & (5, 20) & (-0.9, 1) & 
         0.09 \newline \textit{56\%} \newline 0s  & 0.48 \newline 100\% \newline 0.01s & 0.51 \newline 100\% \newline 1.15s& \textbf{0.35} \newline 100\% \newline 0.03s  & 0.40 \newline 100\% \newline 1.89s\\
  \hline       

         [5, 5, 5, 5], \newline  [20, 50, 2, 8] & (15, 60) & (0, 0.6) & 
         0.4 \newline  96\% \newline 0s  & 0.54 \newline 100\% \newline 0.02s & 0.47 \newline 100\% \newline 13.24s & 0.51 \newline 100\% \newline 1.04s  & \textbf{0.40} \newline 99\% \newline 438.26s\\
  \hline       

         [2, 12, 0, 1], \newline [2, 55, 1, 2]& (10, 40) & (0.8, 0.9) & 0.22 \newline  95\% \newline 0s  & 0.35 \newline 100\% \newline 0.02s & 0.28 \newline 100\% \newline 1.04s&0.31 \newline 99\% \newline 0.23s  & \textbf{0.22} \newline 98\% \newline 43.41s\\
  \hline       

         [2, 2, 12, 4], \newline  [3, 64, 1, 2] & (5, 60) & (-0.5, 0.9) & 0.28 \newline  97\% \newline 0s  & 0.59 \newline 100\% \newline 0.01s & 0.53 \newline 100\% \newline 4.66s&0.54 \newline 100\% \newline 0.16s  & \textbf{0.44} \newline 100\% \newline 60.12s\\
  \hline       

         [0, 16, 0, 4], \newline  [3, 9, 1, 7], \newline [5, 5, 5, 5] & (5, 10, 15) & (0.8, 0.4, 0) &
         0.49 \newline  98\% \newline 0s  & 0.76 \newline  100\% \newline  0.01s & 0.64 \newline  100\% \newline 287.23s&0.65  \newline  100\% \newline 0.08s  & \textbf{0.51} \newline  100\%  \newline 1277.17s\\
  \hline       

         [1, 13, 1, 0],  \newline  [0, 18, 0, 2],
          \newline  [0, 20, 0, 5] & (10, 10, 10) & (0.8, 0.9, 0.8) & 
         0.3 \newline  97\%  \newline  0s  & 0.51 \newline 100\%  \newline  0.01s & 0.36  \newline  100\%  \newline 59.4s& 0.43  \newline  100\%  \newline  0.09s  & \textbf{0.29}  \newline  99\% \newline 1729.77s\\
  \hline       

         [0, 19, 1, 0], \newline  [3, 4, 4, 4], \newline [0, 2, 18, 0] & (5, 5, 5) & (0.9, 0, -0.8) & 0.41 \newline  99\% \newline 0s  & 0.81 \newline 100\% \newline 0.01s & 0.65 \newline 100\% \newline 244.09s& 0.69 \newline 100\% \newline 0.07s  & \textbf{0.55} \newline 100\% \newline 81.74s\\
  \hline       

         [5, 0, 0, 5], \newline  [6, 0, 0, 14], \newline [18, 1, 1, 10] & (5, 5, 25) & (0, 0, 0) & 0.61 \newline \textit{92\%} \newline 0s  & 0.88 \newline 100\% \newline 0.01s & 0.77 \newline 100\% \newline 197.29s&0.74 \newline 100\% \newline 0.09s  & \textbf{0.58} \newline 98\% \newline 469.31s\\
    \bottomrule
    \end{tabular}%
    }
    \caption{ Average width, empirical coverage rates, and mean runtimes for confidence intervals for $\tau$ in unbalanced experiments. 
    The three rows in the last 5 columns report average width, empirical coverage rate, and average runtime, respectively, in 100 replications. 
    ``W-S'' is the difference in Bonferroni-adjusted Wendell-Schmee intervals.
    ``ESI'' is the difference in Bonferroni-adjusted ``exact stratified inference'' intervals. 
    ``CPT'' is the combined stratumwise permutation tests and ``SPT'' is the stratified permutation test.  
    Tests in the last two columns were based on 100~pseudorandom treatment assignments.
    Scenarios in which the empirical coverage was less than the nominal level of 95\% are italicized.
    The narrowest mean width among the conservative methods is in bold font.}

    
    \label{tab_simu}
\end{table}

\begin{table}
    \centering
    \small
    \resizebox{\columnwidth}{!}{%
    \begin{tabular}{p{2.1cm}p{2.5cm}p{2.1cm}|p{1cm}p{1cm}p{1cm}}
    \toprule
         $\bv$ & $\bn$ & $\tau$ &  \textbf{Wald} & \textbf{ESI}  & \textbf{CPT} 
         \\
         \hline
         [8, 15, 0, 7], 
         \newline  
         [9, 21, 1, 9], 
         \newline 
         [12, 26, 1, 11] & (10, 10, 10) & (0.5, 0.5, 0.5) & 0.36 
         \newline  
         97\% 
         \newline 
         0s  & 0.55 
         \newline 
         99\% 
         \newline 
         0.02s & \textbf{0.48} 
         \newline 
         99\% 
         \newline 
         0.6s  
         \\
  \hline       

         [10, 20, 0, 10], \newline  [7, 1, 25, 7], \newline [12, 8, 8, 12] &(20, 30, 10) & (0.5, -0.6, 0) & 0.36 \newline  97\% \newline 0s  & 0.58, \newline 100\% \newline 0.02s & \textbf{0.51} \newline 100\% \newline 0.53s  
         \\
  \hline       

         [5, 0, 0, 45], \newline  [10, 0, 0, 40], \newline [10, 0, 0, 40] \newline [5, 0, 0, 75] & 
         (25, 25, 25, 40) & 
         (0, 0, 0, 0)& 0.17 \newline  \textit{91\%} \newline 0s  & \textbf{0.29} \newline 100\% \newline 0.05s &0.34, \newline 100\% \newline 6.11s  
         \\
\hline
        [15, 8, 0, 7],\newline [9, 1, 21, 9],\newline[7, 6, 21, 6],\newline [3, 12, 3, 3],\newline[5, 20, 1, 1]&(10, 10, 10, 10, 10) & (0.3, -0.5, 0.4, 0.4, 0.7) &
        0.4
        \newline 97\%
        \newline 0s &
        0.56\newline 100\%\newline 0.03s & \textbf{0.52}\newline 100\%\newline 14.96s
        \\
\hline
        [1, 0, 0, 39],\newline [10, 0, 0, 40],\newline [10, 0, 0, 10],\newline [3, 0, 0, 7],\newline [25, 0, 0, 25]&(20, 30, 10, 5, 20) &
        (0, 0, 0, 0, 0) &
        0.25\newline 98\% \newline 0s & \textbf{0.47}\newline 100\% \newline 0.03s &0.48\newline 100\% \newline 9.73s
        \\
         \bottomrule
    \end{tabular}%
    }
    \caption{ Average width, empirical coverage rates, and mean runtimes for confidence intervals for $\tau$ in unbalanced experiments. 
    The three rows in the last three columns report average width, empirical coverage rate, and average runtime, respectively, in 100 replications. 
    ``ESI'' is the difference in Bonferroni-adjusted ``exact stratified inference'' intervals. 
    ``CPT'' is the combined stratumwise permutation test.
    Tests in the last column were based on 100~pseudorandom treatment assignments.
    The Wendell-Schmee (W-S) and stratified permutation test (SPT) methods did not complete within 30 minutes in these examples;
    no results are listed for them.
    Scenarios in which the empirical coverage rate of Wald's interval was below 95\% are italicized.
    The narrowest mean width among the two conservative methods is in bold font.}

    
    \label{tab_simu_large}
\end{table}

We ran an additional simulation study of ``balanced'' 
experiments: 
in each stratum, the number of subjects assigned to treatment is equal to the number assigned to control.
Simulations were carried out for two strata and various values of $n$, $\tau_1$, and $\tau_2$,
as follows.
The value of $\tau_k$ is always a multiple of $1/n_k$, so $\tau_kn_k$ is an integer:
\begin{enumerate}

    \item 
    For $r=1, \ldots, 100$:
    \begin{itemize}
        \item For $k = 1, \ldots, K$, 
        set $y_{kj}(1) \leftarrow 1$ and $y_{kj}(0) \leftarrow 0$ for subjects $j=1, \ldots, \tau_k n_k$. 
        For $j= \tau_{k} n_{k}+1, \ldots, n_k$, set $y_{kj}(1)$ to be
        a draw from a Bernoulli($1/2$) distribution.
        Let $y_{\tau_{i}n_{k}+1}(0), \ldots, y_{n_k}(0)$ be a pseudorandom permutation of 
        $\{ y_{\tau_{k}n_{k}+1}(1),\ldots,y_{n_k}(1)\}$, 
        so the 
        stratum ATE is $\tau_k$.
        \item Generate data for stratum $k$ by pseudorandomly assigning $n_k/2$ subjects to treatment and the rest to control.
        \item Compute 95\% (nominal) confidence intervals using all five methods.
        (The permutation tests used
        100~pseudorandom treatment allocations.) 
    \end{itemize}
    
    \item Calculate the average width of the confidence intervals for each method across the 100 replications.
\end{enumerate}
\begin{figure}
    \centering
    \includegraphics[width=1\linewidth]{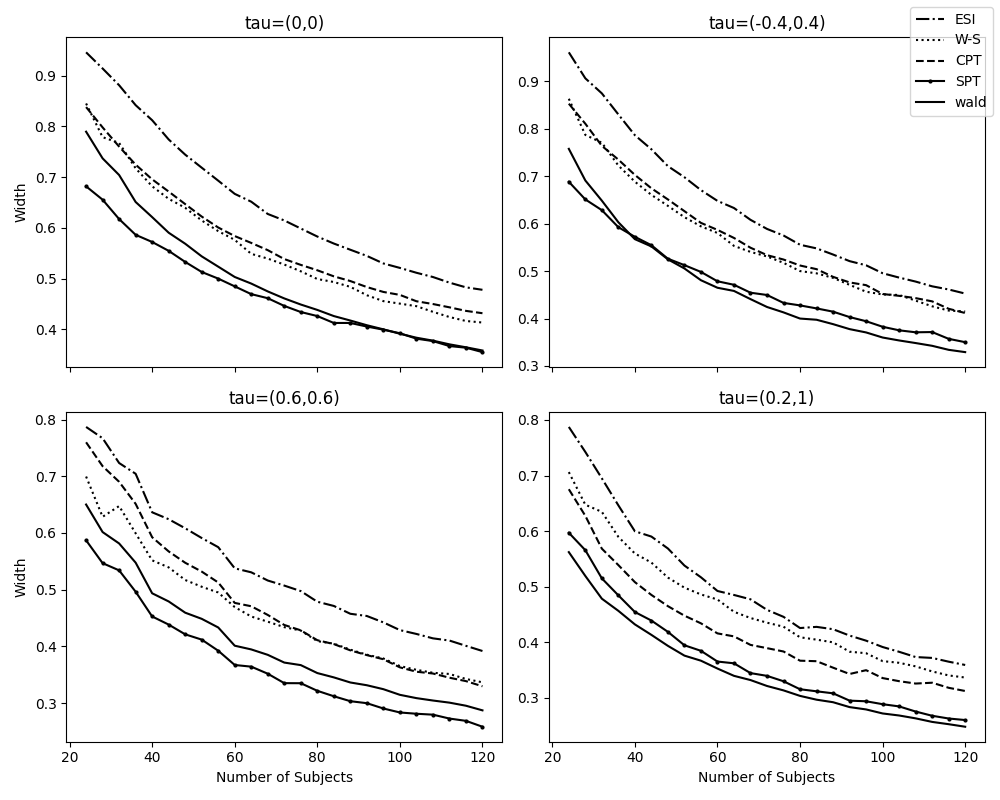}
    \captionsetup{font=small}

    \caption{Average widths of (nominal) 95\% confidence intervals 
    for five methods in balanced experiments with different numbers of subjects. 
    X-axis: number of subjects. 
    Y-axis: mean width in 100
    replications.
    The CPT and SPT methods used 100~pseudorandom treatment allocations to find each confidence set.
    The Wald intervals are not guaranteed to attain their nominal coverage level; nonetheless, they are not always shortest.
    \label{fig:simu}}
\end{figure}
Figure~\ref{fig:simu} shows that 
among the four conservative methods, SPT produces the shortest intervals on average, followed by W-S and CPT.
ESI produced the widest intervals on average.
The intervals based on Wald's method fail to have their nominal coverage (the coverage is as low as 56\% in one of the examples); and they are not always the shortest in these examples 
(their average length is between those of SPT and W-S).

\subsection{Case study: vedolizumab for chronic pouchitis}
We apply the methods to a recent
stratified experiment on
the efficacy of vedolizumab 
to treat chronic pouchitis \citep{travis_casestudy_2023}.
Chronic pouchitis (long-term inflammation of the ileal pouch associated with increased stool frequency, abdominal pain, and impaired quality of life) is the most common complication among patients who undergo treatment for ulcerative colitis---inflammation and ulcers in the colon and rectum.  
Vedolizumab is a monoclonal antibody approved to treat moderately to severely active ulcerative colitis and Crohn's disease in adults, with potential  as an off-label treatment for chronic pouchitis.
The experiment randomized 102~patients to receive vedolizumab or placebo, stratified according to whether the subject was  taking antibiotics continuously at baseline.
The binary outcome measure was remission, defined as reduction of the
modified Pouchitis Disease Activity Index (mPDAI) by at least two points at week~14 \citep{travis_casestudy_2023}.
In the stratum that received continuous antibiotics, the remission rate was 27.6\% (8 of 29 patients) among the treated and 12.0\% (3 of 25 patients) in the placebo group.
In the other stratum, the remission rate was 36.4\% (8 of 22 patients) among the treated and 7.7\% (2 of 26 patients) in the placebo group. 
Table~\ref{tab_case_study} gives the 95\% confidence intervals derived from the four new methods, the Wald method, and the Cochran-Mantel-Haenszel (CMH) interval reported by \cite{travis_casestudy_2023}.

\begin{table}
    \centering
    \begin{tabular}{ccc}\hline
         \textbf{Method} & \textbf{95\% CI} & \textbf{Width}\\
         \hline
         W-S & [0.04, 0.39] & 0.35\\
         ESI & [-0.01, 0.41] & 0.42\\
         SPT & [0.06, 0.38] & 0.32\\
         CPT & [0.02, 0.40] & 0.38 \\
         Wald & [0.06, 0.38] & 0.32 \\
         CMH & [0.05, 0.38] & 0.33\\
         \hline
    \end{tabular}
    \caption{Confidence intervals for the effect of vedolizumab on chronic pouchitis \citep{travis_casestudy_2023}
    using five methods,
    of which three are conservative (W-S: combining Bonferroni-adjusted Wendell-Schmee intervals;
    ESI: combining Bonferroni-adjusted ``exact stratified inference'' intervals; SPT: stratified permutation test; CPT: combined stratumwise permutation tests) and two are approximate (Wald; CMH: Cochran-Mantel-Haenszel, the interval reported in the original study).
    Among the conservative methods, SPT gives the shortest interval---which happens to be identical CMH interval in this example.}
    \label{tab_case_study}
\end{table}
In this example, the SPT confidence set is the narrowest of the four exact approaches.
It coincides with the Wald interval; both are slightly narrower than the CMH interval.
However, unlike the Wald or CMH intervals, which may undercover, the confidence level of the SPT interval really is 95\%.

\section{Extensions and Directions for Improvement}

The methods discussed here can be extended to handle missing data and to make inferences about relative risk, which some researchers consider more interpretable than ATE. 
They can also be extended to other test statistics.

\subsection{Missing Data}
In real experiments, data are 
often missing for a subset of subjects.
Missing observations can be handled by 
substituting the most conservative value for each endpoint of the confidence interval:
To compute the upper confidence bound on the treatment effect,
treat missing control cases as 0 and missing treated cases as 1.
To compute the lower confidence bound, do the opposite.
This approach is computationally tractable
whenever the full-data
problem is tractable.
Using multiple imputation or other model-based approaches to fill in missing data may undermine the conservative nature of the approaches presented here.
However, see \cite{ivanovaEtal22}.

\subsection{Relative Risk}
The approaches in Section~\ref{sec:exact_ci} also work to find confidence intervals for relative risk (RR),
the fraction of responses equal to 1 if all subjects had been assigned to the active treatment,
$v_{1\bullet}/n$,
divided by the fraction of 
responses equal to 1 if all subjects 
had been assigned to control,
$v_{\bullet1}/n$.
One can find a confidence interval for RR inexpensively by computing
simultaneous $1-\alpha$ confidence
intervals for $v_{1 \bullet}$
and $v_{\bullet 1}$,
for instance 
by finding level $1-\alpha/2$
confidence intervals 
using the methods outlined in Section~\ref{sec:hyper_stratified},
then taking their ratio.
A sharper approach is to find the smallest and largest values of RR in a confidence set for the potential outcome table derived using methods in Section~\ref{sec:invert_permute}.

While confidence bounds for RR can be based on any test statistic, different statistics generally yield different confidence intervals.
For instance, basing the test on the absolute difference between the estimated and true RR might yield shorter intervals on average than basing the test on the absolute difference between the estimated and true ATE.

However, if the same test statistic and the same set of permutations is used to construct confidence intervals for ATE and for RR, 
the confidence bounds for the two quantities have simultaneous coverage probability: whenever the confidence interval for the ATE covers, 
the confidence interval for RR also covers.
Such a ``strict bounds'' approach \citep{evansStark02} makes it possible to construct simultaneously valid confidence intervals for any number of parameters without the need to adjust for multiplicity.

\subsection{Other test statistics}

Appendix~\ref{sec:other_test_stat} shows that Sterne's method \citep{sterne_1954}
applied to the difference between the estimated and true ATE may yield shorter confidence intervals on average than inverting tests based on the absolute difference between the estimated and true ATE.
However, finding the acceptance regions for Sterne's method involves stochastic approximation (so the resulting intervals are not guaranteed to be conservative) and is computationally expensive.
Developing a conservative, computationally tractable implementation of Sterne's method---perhaps using ideas from exact inference via Monte Carlo methods---may improve the statistical efficiency of inferences.

\section{Conclusions}
We introduced four methods for constructing exact or conservative finite-sample confidence sets for the ATE in stratified randomized experiments with binary outcomes.
The stratified permutation test (SPT) extends the methods developed by  \citet{rigdon_binary_2015} and \citet{li_binary_2016} to stratified experiments. 
The combined permutation test (CPT) 
uses $P$-value combining functions to pool evidence across strata,
decomposing a hypothesis about the ATE into a union of intersections of hypotheses about the ATE in each stratum separately.
The methods based on ``efficient stratified inference'' (ESI) \citet{stark19} and \citet{wendell_strata_1996} (W-S)
are natural generalizations of methods for 
finding confidence bounds for the ATE in unstratified experiments using Bonferroni-adjusted hypergeometric confidence sets.

The coverage of Wald's method was below its nominal level when 
the number of subjects in treatment and control was not balanced, 
when the ATE was close to $1$ or $-1$ in every stratum, 
and when treatment had no effect on any individual, i.e.,
when the strong null hypothesis was true.
Wald intervals were not always the shortest, despite their sometimes inadequate coverage.
Among the conservative methods, SPT generally yields the narrowest confidence intervals, followed by W-S and CPT,
but SPT and W-S are computationally intractable when there are more than three strata.
ESI generally gave the
widest confidence intervals, but it sometimes gave the shortest intervals among the conservative methods that are computationally tractable
for four or more strata

There is a tradeoff between computational efficiency and statistical efficiency:
ESI is computationally tractable for arbitrarily many strata and for large strata.
CPT requires roughly one or two orders of magnitude more CPU time in our (small) examples: $K \le 4$ and $n \le 230$.
With current personal computers, W-S and SPT are limited to experiments with few strata ($K \leq 3$) and relatively few subjects ($n \le 100$).

Promising directions for future research include extending these methods from binary responses to bounded responses,
extending the methods to accommodate more than two treatment levels,
finding a multiplicity adjustment (for W-S and ESI) that is sharper than Bonferroni's by taking into account the dependence explicitly, and improving the computational efficiency of SPT to make it tractable for larger experiments. 

\bibliography{stratified_binary_ate}
\newpage
\appendix

\section{Additional Detail}
\label{sec:addition}

\subsection{Other test statistics}
\label{sec:other_test_stat}

In the methods that used permutation tests (including the unstratified method in Section~\ref{sec:inverting_unstratified}, SPT and CPT) we generally used the absolute difference $|\hat{\tau}(\bn) - \tau(\bv)|$ as the test statistic.
We tried two other test statistics: the studentized 
absolute difference
between the sample ATE and the true ATE,
and Sterne's method \citep{sterne_1954} for the signed estimated ATE, $\hat{\tau}(\bn)$, which uses the smallest acceptance region for each null rather than regions that are symmetric around the null. 

The studentized absolute difference is $|\hat{\tau}(\bn) - \tau(\bv)|/\widehat{V}(\bn)^{1/2}$, where $\widehat{V}(\bn)$ is the estimated variance of $\hat{\tau}$ (see (\ref{equ_wald_ci})):

\[
   \widehat{V}(\bn)=
  \sum_{k=1}^{K} \left ( \frac{n_k}{n} \right )^2
        \left (
        \frac{\widehat S^2_k(1)}{m_k}+\frac{ \widehat S^2_k(0)}{n_k-m_k}
        \right ).
\]
Table~\ref{tab_test_stat}
compares confidence intervals based on the raw and studentized absolute difference between the sample ATE and the true ATE.
Table~\ref{tab_test_stat} suggests that the absolute difference statistic performs better,
especially when the experiment is unbalanced. 
This could be because,
for some potential outcome tables, 
the variability of $\widehat{V}(\bn)$ increases the variability of the test statistic.
Moreover, it is possible that  the estimated variance is $0$, making the statistic infinite. 

         











To use Sterne's method, 
we first identified the distribution of $\hat{\tau}(\bbn(\bv,\bZ))$. 
We set the acceptance region for Sterne's method empirically by finding the narrowest range of values of $\hat{\tau}$ that contained $(1-\alpha)\times 100\%$  of the samples.
Table~\ref{tab_test_stat} shows the confidence intervals for the ATE for various observed tables, along with the number of potential outcome tables that were rejected for each observed table. 
Sterne's method yields the shortest intervals more often than the other methods. 
However, our implementation of Sterne's method involves approximating the acceptance regions by simulation, so its coverage is not guaranteed; moreover, it is much more expensive to compute than the absolute difference. 
Until there is an implementation that is computationally efficient and guaranteed to be conservative, we recommend  
using the absolute difference between the sample ATE and the true ATE as the test statistic, as described in 
Section~\ref{sec:inverting_unstratified}.

\begin{table}
    \centering
    \small
    \begin{tabular}{lcrcrcr}
    \toprule
         $\bn$ & 
         \multicolumn{2}{c}{\textbf{\parbox{2cm}{Absolute difference}}}
         &
          \multicolumn{2}{c}{\textbf{\parbox{2cm}{Studentized absolute difference}}}
         &
         \multicolumn{2}{c}{\textbf{Sterne's method}}\\
         \cmidrule(lr){2-3} \cmidrule(lr){4-5} \cmidrule(lr){6-7}
         & \textbf{interval} &
         \textbf{length} &
         \textbf{interval} &
         \textbf{length} &
         \textbf{interval} &
         \textbf{length}\\
         \midrule
         
         [10, 10, 10, 10]& [-0.28, 0.28] & 0.56 & [-0.30,0.30] &0.60&[-0.25, 0.25] & \textbf{0.50}
         \\

         [10, 10, 9, 10]& [-0.26, 0.31] &  \textbf{0.56} &[-0.26,0.33]& 0.59 & [-0.26, 0.31] &   \textbf{0.56}\\

         [10, 10, 1, 19]& [0.18, 0.65]& 0.47 &[0.23,0.65]& \textbf{0.42}&[0.20, 0.63] & 0.43\\

         [19, 1, 1, 19]& [0.73,  0.93]& 0.20 &[0.75,0.93]& \textbf{0.18}&[0.75, 0.93] & \textbf{0.18}\\

         [6, 6, 10, 10]& [-0.28, 0.28]& \textbf{0.56} &[-0.34,0.34]& 0.68&
         [-0.31, 0.31] & 0.62\\

         [6, 6, 1, 19]& 
         [0.16, 0.66]& \textbf{0.50} 
         &[0.13,0.75]& 0.62&
         [0.16, 0.69]& 0.53\\

         [11, 1, 1, 19]& [0.63, 0.90]& 0.27 &[0.66,0.90]& 0.24 &[0.81, 0.97]& \textbf{0.16}\\

         [2, 2, 15, 15]& 
         [-0.38, 0.38] & \textbf{0.76} &[-0.44,0.44]&0.88 &
         [-0.38, 0.38]& \textbf{0.76}\\

         [2, 2, 2, 28]& 
         [0, 0.82]& \textbf{0.82}&[-0.06,0.85]& 0.91&
         [0, 0.82]& \textbf{0.82}\\

         [3, 1, 2, 28]& 
         [0.24, 0.88]& 0.64 & [0,0.88] & 0.88 &
         [0.50, 0.97]& \textbf{0.47}\\

         \bottomrule
    \end{tabular}
    \caption{Confidence intervals for a variety of observed tables in unstratified experiments using
    different test functions. 
    Bold font indicates the method that produced the narrowest confidence intervals. 
    Absolute difference is $|\hat{\tau} - \tau|$.
    Studentized absolute difference is the absolute difference divided by an estimate of the standard error
    of $\hat{\tau}$.
    Sterne's method uses the narrowest acceptance region for each null,
    approximated by simulation.
    Sterne's method yields the shortest intervals most often but are approximate and the most expensive to compute.}
    \label{tab_test_stat}
\end{table}

\subsection{Other combining functions}
\label{sec:combining_choice}


Section~\ref{sec:invert_permute_combine} introduced CPT using Fisher's combining function. 
This section uses simulation to compare 
the performance of CPT with Fisher's combining function to CPT with other common $P$-value combining functions: 
Pearson's method, Mudholkar and George's method, Tippett's method, and Stouffer's Z-score, all of which are available in 
the {\tt{scipy.stats}} Python package \citep{2020SciPy-NMeth}. 


The simulations used the potential outcome tables in Table~\ref{tab_simu}.
For each table, we generated 100 sets of observed data and calculated the average width of the 95\% confidence intervals obtained using the combining permutation method with various combining functions,
along with the generated observed data. 
The Monte Carlo permutation tests, summarized in Table~\ref{tab_simu_comb}, used 
a random sample with replacement of 100 treatment allocations.
Fisher's combining function generally yielded the narrowest intervals, typically followed and occasionally bested by Tippett's combining function.
Tippett's combining function saves computation
and is the only method that beat Fisher's combining function in any of the experiments.

\begin{table}
    \centering
    \small
     \resizebox{\columnwidth}{!}{%
    \begin{tabular}{p{2.2cm}p{1.2cm}p{1.7cm}|p{1cm}p{1cm}p{1cm}p{1cm}p{1cm}}
    \toprule
         $\bv$& $\bn$ & {$\bf{\tau}$} &   \textbf{Fisher} & \textbf{Pearson} & \textbf{George} & \textbf{Tippett} & \textbf{Stouffer}\\
         \hline
         
        [10, 10, 10, 10], \newline  [10, 10, 10, 10] & (10,10) & (0,0) &\textbf{0.57}& 0.74& 0.59&0.67 & 0.6\\
        \hline
         
         [3, 8, 4, 5], \newline  [0, 19, 1, 0] & (15,15) & (0.2,0.9) & \textbf{0.65}&0.84&0.73&0.7&0.74\\
         \hline

         [3, 23, 2, 2], \newline  [4, 2, 30, 4] & (5,30) & (0.7,-0.7) & 0.61&0.93&0.86&\textbf{0.59}&0.87\\
         \hline

         [2, 24, 0, 4], \newline  [1, 26, 2, 1] & (5,25 )& (0.8,0.8) & \textbf{0.53}&0.71&0.61&0.62&0.62\\
         \hline

         [1, 0, 9, 0], \newline  [0, 40, 0, 0] & (5,20) & (-0.9,1) & 0.36&1&1&\textbf{0.27}&1\\
         \hline

         [5, 5, 5, 5], \newline  [20, 50, 2, 8] & (15,60) & (0,0.6) & \textbf{0.51}&0.9&0.83&0.52&0.84\\
         \hline
         
         [2, 12, 0, 1], \newline  [2, 55, 1, 2] & (10,40) & (0.8,0.9) & \textbf{0.32}&0.87&0.83&0.34&0.84\\
         \hline

          [2, 2, 12, 4], \newline  [3, 64, 1, 2] & (5,60) & (-0.5,0.9) &0.56&0.97&0.93&\textbf{0.5}&0.93\\
          \hline

          [0, 16, 0, 4], \newline  [3, 9, 1, 7], \newline [5, 5, 5, 5] & (5,10,15) & (0.8,0.4,0) &\textbf{0.66}&0.87&0.79&0.79&0.79\\
          \hline

         [1, 13, 1, 0], \newline  [0, 18, 0, 2], \newline [0, 20, 0, 5] &(10,10,10) & (0.8,0.9,0.8) & \textbf{0.42}&0.81&0.68&0.56&0.68\\
         \hline

          [0, 19, 1, 0], \newline  [3, 4, 4, 4], \newline [0, 2, 18, 0]  & (5,5,5) & (0.9,0,-0.8) &\textbf{0.69}&0.98&0.96&0.71&0.96\\
          \hline

          [5, 0, 0, 5], \newline  [6, 0, 0, 14], \newline [18, 1, 1, 10] & (5,5,25) & (0,0,0) &\textbf{0.74}&0.91&0.86&0.85&0.86\\
          \hline

         [8, 15, 0, 7], \newline  [9, 21, 1, 9], \newline [12, 26, 1, 11]& (10,10,10) & (0.5,0.5,0.5) & \textbf{0.47}&0.8&0.67&0.64&0.67 \\
         \hline

         [10, 20, 0, 10], \newline  [7, 1, 25, 7], \newline [12, 8, 8, 12]& (20,30,10) & (0.5,-0.6,0) & \textbf{0.52}&0.9&0.78&0.64&0.78\\
         \hline

          [5, 0, 0, 45], \newline  [10, 0, 0, 40], \newline [10, 0, 0, 40] \newline [5, 0, 0, 75] & (25,25, \newline  25,40)& (0,0,0,0)&\textbf{0.34}&0.83&0.63&0.49&0.64\\
         
         \bottomrule
    \end{tabular}
    }
    
    \caption{Simulation results for different potential outcome tables. 
        Average width of 95\% confidence intervals
        in 100 replications using the
        CPT with different $P$-value combining functions. 
        Bold font indicates the narrowest confidence interval on average in each scenario.}
    \label{tab_simu_comb}
\end{table}

To obtain a confidence interval using Tippett's method does not require testing every combination of stratumwise treatment effects, resulting in computational savings.
Instead, one can first find individual $(1-\alpha)^{1/K}$ confidence intervals for the ATE in each stratum, then, find a weighted linear combination of the endpoints to create the confidence interval for the whole study. 
This is equivalent to combining the stratumwise confidence intervals using inverted permutation tests with \v{S}id\'{a}k's correction:
the $P$-value for Tippett's method is 
\[
1-(1-\min(p_1,p_2,.\ldots,p_K))^{K}.
\]

\section{Reducing the Computational Cost of SPT}
\label{sec:reduce_time}

Suppose we have an observed table $\bn$. 
If there is a stratum $k$ for which $n_k = 2 m_k$, we say the experiment is \emph{balanced in stratum} $k$.
In this section, we suppose that the first $\ell \ge 1$ strata are balanced. 
We will show that then we only need to test 
$O((\sum_{k=1}^\ell n_k)\times\prod_{k=1}^\ell n_k^2\times\prod_{k=\ell+1}^{K} n_k^{3})$ 
potential outcome tables. 
If the strata have similar sizes, that number is about $O(n^{3K-\ell+1})$, 
much smaller than the number required by
the approach introduced in Section~\ref{sec:exact_ci}, which is $O(n^{3K})$.
\begin{lemma}
\label{lem_hattau_x_partial_balanced}
Consider a potential outcome table $\bv=(\bv_1,\bv_2, \ldots, \bv_K)^T$ and
realized treatment assignment vector $\bz$, where the first $\ell \ge 1$ strata are balanced. 
Let 
$$
x_{kab} := \#\{c \in [n_k] : y_{kc}(1) = a,\; y_{kc}(0) = b,\; z_{kc} = 1\}.
$$
That is, $x_{kab}$ is the number of subjects in stratum $k$ whose response if assigned to treatment is $a$, whose response if assigned to control is $b$, and who were in fact assigned to treatment.
Then, 
\begin{align*}
    \hat{\tau}(\bv,\bz)={}&
    \frac{\sum_{k=1}^\ell (2x_{k11}-v_{k11})}{n}
    -\frac{\sum_{k=1}^\ell(2x_{k00}-v_{k00})}{n} + \\
    {}&\frac{\sum_{k=1}^\ell (v_{k10}-v_{k01})}{n}
    +\frac{1}{n}\sum_{k=\ell+1}^{K}n_k\hat{\tau}_k.
\end{align*}
\end{lemma}
\begin{proof}

The number of subjects in the treatment group whose response is 1 is
$(x_{k11}+x_{k10})$; the corresponding number for the control group is
$(v_{k11}-x_{k11}+v_{k01}-x_{k01})$.
Thus
    \begin{eqnarray*}
        \hat{\tau}(\bv,\bz)-\frac{1}{n}\sum_{k=\ell+1}^{K} n_k \hat{\tau}_k
         & =&\frac{1}{n}\sum_{k=1}^\ell n_k \times \hat{\tau}_k(\bY, \bZ) \\
        &=& 
        \frac{1}{n}\sum_{k=1}^\ell n_k \times \left(\frac{x_{k11}+x_{k10}}{m_k} \right. \\
        && 
        \;\;\;\;\;~~~~
        -\left.\frac{ v_{k11}-x_{k11}+v_{k01}-x_{k01}}{n_k-m_k} \right ) \\
        &=&
        \frac{2}{n}\sum_{k=1}^\ell [x_{k11}+x_{k10} - (v_{k11}-x_{k11}+v_{k01}-x_{k01})].
    \end{eqnarray*}
The last equation holds since for $k \le \ell$, $n_k = 2m_k = 2(n_k-m_k)$. 
Now $\forall k$,
\[
    x_{k11} + x_{k10}+x_{k00}+x_{k01} = m_k,
\]
so $\forall k \le \ell$,
\[
     (v_{k11} + v_{k10} + v_{k00}+v_{k01}) - (x_{k11} + x_{k10} + x_{k00} + x_{k01}) = n_k - m_k = m_k.
\]
Plugging in
$x_{k11} + x_{k10} = m_k - x_{k00} - x_{k01}$ 
and
$v_{k11}-x_{k11}+v_{k01}-x_{k01} = m_k - (v_{k00} - x_{k00} - v_{k10} - x_{k10})$
we have
    \begin{eqnarray*}
        \hat{\tau}(\bv,\bz)-\frac{1}{n}\sum_{k=\ell+1}^{K} n_k \hat{\tau_k}
        &=&\frac{1}{n} \left [ \sum_{k=1}^\ell
        (x_{k11}+x_{k10}) + \sum_{k=1}^\ell(m_k-x_{k00}-x_{k01}) \right ] - \\
        && - \frac{1}{n} \left [ \sum_{k=1}^\ell (v_{k11}-x_{k11}+v_{k01}-x_{k01}) \right .
        - \\
        && \;\;\;\;\; - \left . \sum_{k=1}^\ell (m_k-(v_{k00}-x_{k00}+v_{k10}-x_{k10})) \right ]. 
\end{eqnarray*}
Simplifying this last equation yields
\begin{eqnarray*}
        \hat{\tau}(\bv,\bz)-\frac{1}{n} \sum_{k=\ell+1}^{K} n_k \hat{\tau_k}= 
        \frac{1}{n} \sum_{k=1}^\ell \left [ (v_{k10}-v_{k01}) +  (2x_{k11}-v_{k11}) -(2x_{k00}-v_{k00})
        \right ].
\end{eqnarray*}
\end{proof}
\Cref{lem_hattau_x_partial_balanced} shows that the distribution of $\hat{\tau}(\bv,\bz)$ 
depends on $v_{k10}$ and $v_{k01}$ in the balanced strata
only through $\sum_{k=1}^\ell (v_{k10}-v_{k01})$. 
The following theorem uses that fact to characterize sets of potential outcome tables that produce the same $P$-value as each other under the same randomization. 

\begin{theorem}
    \label{thm:same_p_partial_balanced}
    Consider an observed table $\bn$ in which the first $\ell$ strata are balanced. 
    If two potential outcome tables $\bv$ and $\bv'$ satisfy
    \begin{enumerate}
        \item $v_{k11}=v'_{k11},\; v_{k00}=v'_{k00},\; \forall k\in [K]$
        \item $v_{k10}=v'_{k10},\; v_{k01}=v'_{k01},\; \forall k \in \{\ell+1,\ell+2, \ldots, K\}$
        \item $\sum_{k=1}^\ell (v_{k10}-v_{k01}) 
        = 
        \sum_{k=1}^\ell (v'_{k10}-v'_{k01})$,
    \end{enumerate}
    then, holding the set of  pseudorandom permutations $\{\bW_r\}_{r=1}^R$ fixed, 
    \[ p(\bv, \bn) = p(\bv',\bn),\]
    where the $P$-value $p(\bv, \bn)$ is defined in (\ref{eq:p-value-def-random}).
\end{theorem}
\begin{proof}
    Recall from \Cref{eqn:ate_potential_outcomes} that 
    $\tau(\bv)=\frac{1}{n}\sum_{k=1}^{K}(v_{k10}-v_{k01})$ for any potential outcome table. 
    Assumptions~2 and 3 then imply that $\tau(\bv)=\tau(\bv')$.
    Furthermore, Lemma~\ref{lem_hattau_x_partial_balanced} implies that 
    $\hat{\tau}(\bv,\bz) = \hat{\tau}(\bv',\bz)$ for any allocation $\bz$ of units to treatment and control in each stratum.
    The desired result follows from the representation of $p(\bv,\bn)$ in (\ref{eq:p-value-def-random}).
\end{proof}

The following algorithm 
applies Theorem~\ref{thm:same_p_partial_balanced} to construct the confidence set described in
Section~\ref{sec:invert_permute}
but examines fewer potential outcome tables.

\begin{algorithm}
    \label{algorithm_partially_balanced}
    
    Input: An observed table $\bn$ and desired confidence level $1-\alpha$
    
    Output: A $1-\alpha$ confidence set $S$ for the ATE $\tau$.
    \begin{enumerate}
        \item Initialize $S \leftarrow \emptyset$, $n_k \leftarrow \sum_{ab} n_{kab}$, 
        $n \leftarrow \sum_k n_k$, $m_k \leftarrow n_{k10} + n_{k11}$, $m \leftarrow \sum_k m_k$.
        \item For every element in the following set: 
        \begin{align*}
            \Big\{
                ( u_{111}, \ldots, u_{K11}, u_{100}, \
                \ldots,& u_{K00}, u_{(\ell+1)10}, \ldots, u_{K10}, a) \Big{|}\\
                &0 \le u_{k11} \le n_{k}, k \in [K], \\
                & 0 \le u_{k00} \le n_{k}-u_{k11},  k \in [K], \\
                & 0 \le u_{k10} \le n_{k}-u_{k11}-u_{k00}, k \in \{\ell+1,\ldots,K\},\\
                & |a|\le \sum_{k=1}^\ell (n_{k}-u_{k00}-u_{k11})
            \Big\},
        \end{align*}
        \begin{enumerate}
            \item Find a potential outcome table $\bv$
            algebraically compatible with
            $\bn$ such that 
            \[
            v_{k11}=u_{k11}, k \in [K]; \;
            v_{k00}=u_{k00}, k \in [K]; \;
            v_{k10}=u_{k10}, k \in \{\ell+1,\ldots,K\}; \;
            \]
            and 
            \[
            \sum_{k=1}^{\ell} (v_{k10}-v_{k01}) = a.
            \]
            (The proof of Theorem~\ref{thm_alg_complex_partial_balanced} shows that this can be done in constant time.)
            If no such algebraically compatible potential outcome table exists, pass.
            \item For the table found in step~(a), if $\tau(\bv) \in S$, pass. 
            Otherwise, 
            perform a permutation test with $\bv$ at level $\alpha$. 
            If $\bv$ is accepted, 
            $S \leftarrow S \cup \tau(\bv)$. 
        \end{enumerate}
        \item Return $S$.
    \end{enumerate}
    
\end{algorithm}

\begin{theorem}
    \label{thm_alg_complex_partial_balanced}
    Algorithm \ref{algorithm_partially_balanced} 
    yields the same $1-\alpha$ confidence set as in 
    section~\ref{sec:invert_permute}
    and tests at most 
    $O((\sum_{k=1}^{l} n_k) \times \prod_{k=1}^{l} n_k^2 \times \prod_{k=l+1}^{K} n_k^3)$ 
    potential outcome tables.
\end{theorem}

The proof involves applying a technical lemma 
to test whether a potential outcome table $\bv$ is algebraically compatible with the observed table $\bn$. 
The proof 
simply invokes the corresponding lemma of \citet{li_binary_2016} for the unstratified case separately to each stratum.

\begin{lemma}
    \label{lem_compatible_table_partial_balanced}
    A potential table $\bv$ is algebraically compatible with the observed table $\bn$ iff for all $k \in [K]$,
    \begin{eqnarray*}
        &&\max \{  2n_{k11}-n_{k}-v_{k11}+v_{k00}, -n_k+v_{k11}+v_{k00}, 
        \\
        && ~~~~~~~n_k-v_{k11}-v_{k00}-2n_{k01}-2n_{k10}, v_{k11}-n_k-v_{k00}+2n_{k00} \}
        \\
        \le && v_{k10}- v_{k01}
        \\
        \le && \min \{  v_{k11}+n_k-v_{k00}-2n_{k01}, n_k+v_{k11}+v_{k00}-2n_{k01}-2n_{k10},
        \\
        && ~~~~~~~n_k-v_{k11}-v_{k00},n_k-v_{k11}+v_{k00}-2n_{k10}
        \}.
    \end{eqnarray*}
    and 
    \[
    0\le v_{k11}\le n_{k01}+n_{k11}\le n_k-v_{k00}\le n_k
    \]
\end{lemma}

\begin{proof}[Proof of Theorem \ref{thm_alg_complex_partial_balanced}]
${}$

\begin{enumerate}

    \item \emph{$S$ is a $1-\alpha$ confidence set.} \\
    This follows from Theorem~\ref{thm:same_p_partial_balanced}.

    \item \emph{The number of permutation tests is as claimed.} 
    \\ This follows from inspection of the algorithm.  
    Algorithm \ref{algorithm_partially_balanced} Step~2 executes the loop 
that performs a permutation test $O((\sum_{k=1}^{\ell} n_k) \times \prod_{k=1}^{\ell} n_k^2 \times \prod_{k=\ell+1}^{K} n_k^3)$ times.

    \item \emph{Other aspects of the algorithm do not dominate the computational burden.}
\\
Step~2(a) of algorithm~\ref{algorithm_partially_balanced}, which does the rest of the work,
has constant runtime: 
Lemma~\ref{lem_compatible_table_partial_balanced}, allows us to check whether a potential outcome table in Step~2(a) is algebraically compatible
with $\bn$ by checking whether $a$ is within the range bounded by the lower and upper limits of $\sum_{k=1}^{\ell}(v_{k10}-v_{k01})$. 
To construct an algebraically compatible potential outcome table, we first identify the smallest integer $ g \in [\ell] $ such that the sum of the lower bounds of $ v_{k10} - v_{k01} $ for strata $ k = 1, \dots, g $ and the upper bounds of $ v_{k10} - v_{k01} $ for strata $ k = g+1, \dots, \ell $ exceeds $ a $. Once such $ g $ is determined, we assign $ v_{k10} - v_{k01} $ to its lower bound for strata $ k = 1, \dots, g $, to its upper bound for strata $ k = g+1, \dots, \ell - 1 $, and set $ v_{\ell 10} - v_{\ell 01} $ to ensure that $ \sum_{k=1}^{\ell}(v_{k10} - v_{k01}) = a $.
\end{enumerate}
\end{proof}

\end{document}